\documentclass{article}
\usepackage{spconf,amsmath,graphicx,amssymb,amsfonts}

\newcommand{\ba}{\begin{array}}
\newcommand{\ea}{\end{array}}
\newcommand{\be}{\begin{displaymath}}
\newcommand{\ee}{\end{displaymath}}
\newcommand{\ben}{\begin{equation}}
\newcommand{\een}{\end{equation}}
\newcommand{\bena}{\begin{eqnarray}}
\newcommand{\eena}{\end{eqnarray}}
\newcommand{\beqa}{\begin{eqnarray*}}
\newcommand{\enqa}{\end{eqnarray*}}

\newcommand{\bc}{\begin{center}}
\newcommand{\ec}{\end{center}}
\newcommand{\bi}{\begin{itemize}}
\newcommand{\ei}{\end{itemize}}
\newcommand{\benu}{\begin{enumerate}}
\newcommand{\eenu}{\end{enumerate}}
\newcommand{\bdes}{\begin{description}}
\newcommand{\edes}{\end{description}}
\newcommand{\bt}{\begin{tabular}}
\newcommand{\et}{\end{tabular}}

\newcommand \abf{{\bf a}}
\newcommand \bbf{{\bf b}}
\newcommand \cbf{{\bf c}}

\newcommand \fbf{{\bf f}}

\newcommand \hbf{{\bf h}}
\newcommand \ibf{{\bf i}}

\newcommand \kbf{{\bf k}}

\newcommand \obf{{\bf o}}

\newcommand \vbf{{\bf v}}
\newcommand \wbf{{\bf w}}
\newcommand \xbf{{\bf x}}

\newcommand \Kbf{{\bf K}}

\newcommand \Wbf{{\bf W}}

\newcommand{\Rset}{{\mathbb R}}
\newcommand{\Cset}{{\mathbb C}}

\newcommand{\circlambda}{\mbox{$\Lambda$
             \kern-.85em\raise1.5ex
             \hbox{$\scriptstyle{\circ}$}}\,}

\usepackage{color}
\usepackage{graphicx}

\makeatletter
\usepackage{cite}
\usepackage{blindtext}
\usepackage{algpseudocode} 
\usepackage{subfig}
\usepackage{graphicx}
\usepackage{bbm}
\usepackage{multirow}
\usepackage{booktabs}
\usepackage{hyperref}

\makeatletter
\newcommand{\printfnsymbol}[1]{%
  \textsuperscript{\@fnsymbol{#1}}%
}
\makeatother

\usepackage{babel}

\title{Multi-Modal Recurrent Fusion for Indoor Localization}

 \name{Jianyuan Yu\sthanks{Equal contribution.}\sthanks{J.Yu is a PhD student at Virginia Tech, Blacksburg, VA 24061, USA.}, Pu~Wang\printfnsymbol{1},  Toshiaki Koike-Akino, Philip V. Orlik}
 \address{Mitsubishi Electric Research Laboratories (MERL),
 Cambridge, MA 02139, USA}

\begin{document}
\ninept
\maketitle

\begin{abstract}
This paper considers indoor localization using multi-modal wireless signals including Wi-Fi, inertial measurement unit (IMU), and ultra-wideband (UWB). By formulating the localization as a multi-modal sequence regression problem, a multi-stream recurrent fusion method is proposed to combine the current hidden state of each modality in the context of recurrent neural networks while accounting for the modality uncertainty which is directly learned from its own immediate past states. The proposed method was evaluated on the large-scale \textit{SPAWC2021 multi-modal localization} dataset and compared with a wide range of baseline methods including the trilateration method, traditional fingerprinting methods, and convolution network-based methods. 
\end{abstract}


\section{Introduction}
WiFi-based indoor localization has received long attraction over the past two decades~\cite{HeChan16, ZafariGkelias19}. Among all frameworks, \emph{fingerprinting-based} methods
provide an efficient solution for online localization with low computational complexity~\cite{BahlPadmanabhan00}. On the other hand, it requires enormous time and resources to construct an offline database with chosen fingerprinting features at locations-of-interest to enable fast online localization.

Existing WiFi-based fingerprinting systems have dominantly used the \emph{coarse-grained} received signal strength indicator (RSSI) and \emph{fine-grained} channel state information (CSI) at sub-$7$ GHz to construct the offline training database as these measurements are easy to access from commodity $802.11 g/n/ac/ax$ Wi-Fi devices \cite{WuXiao13, WangGao15wcnc, WangGao17, ChenZhang17}.~Recently, millimeter-wave (mmWave) channel measurements that account for antenna beamforming, e.g., the \emph{mid-grained} beam SNRs, have been considered for the fingerprinting-based indoor localization \cite{PajovicWang19, WangPajovic19, KoikeWang20, WangKoike20b, YuWang20, YuWang22, KoikeWang22}. Traditional machine learning and advanced deep learning methods have been applied to the Wi-Fi-based fingerprinting data and show promising results~\cite{YoussefAgrawala08, MazuelasBahillo09,  LiZhang14, WangGao16, WangGao17b, HsiehChen19, HoangYuen19}. For instance, the $k$-nearest neighbor ($k$NN), support vector machine (SVM), and decision trees (DT) were applied to the RSSI-based fingerprinting method~\cite{YoussefAgrawala08, WuLi07}. DeepFi exploits $90$ CSI amplitudes from all the subcarriers at all the three antennas using an autoencoder \cite{WangGao15wcnc, WangGao17}. Nevertheless, the Wi-Fi-based methods show high sensitivity to subtle environmental changes and pose the generalization issue caused by the measurement inconsistency over time due to hardware impairments and stochastic perturbations.

\begin{figure}[t]
     \begin{center}
     \includegraphics[width=0.46\textwidth]{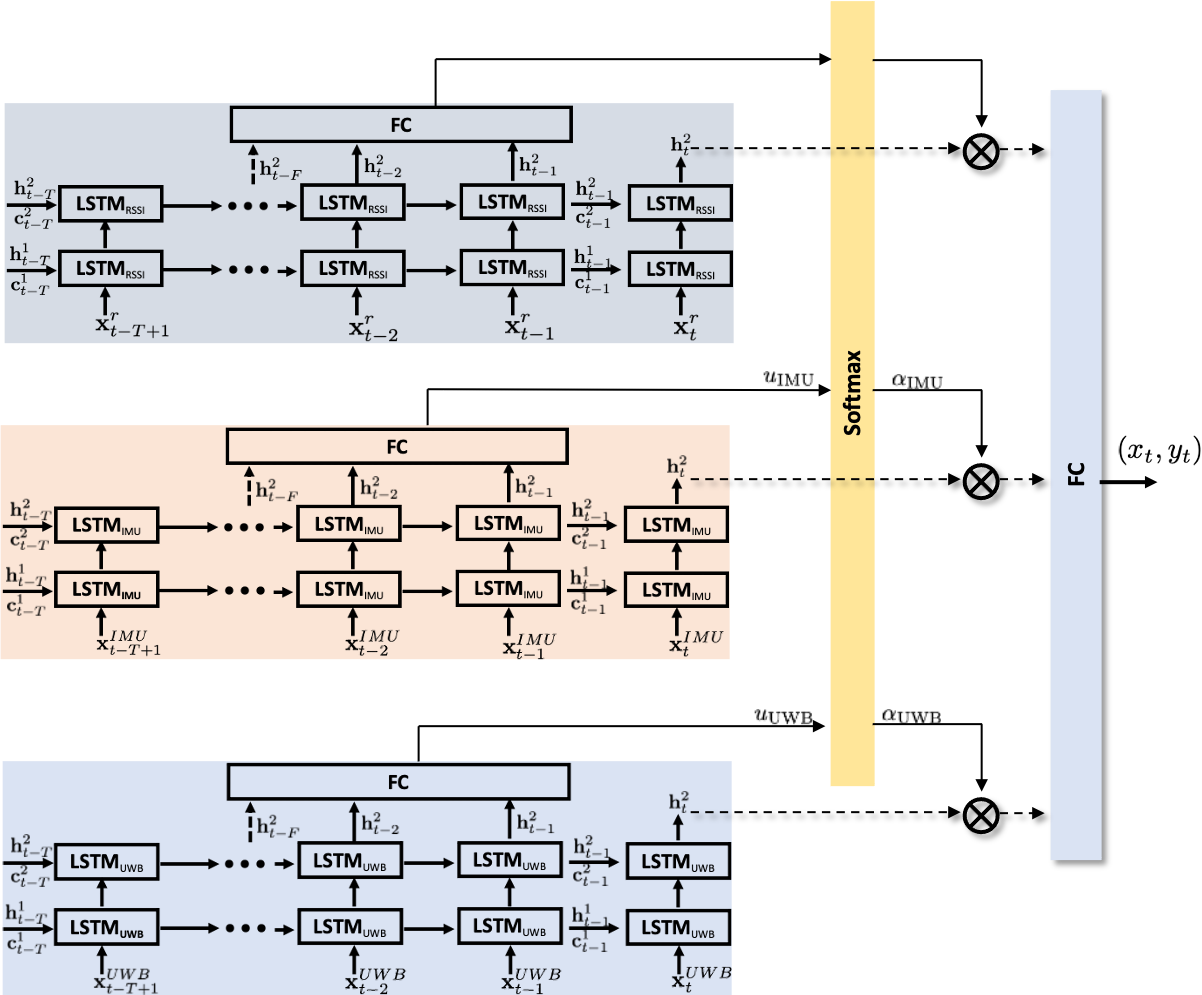}
        \end{center}
           \vspace{-0.15in}
    \caption{Multi-modal recurrent fusion network for indoor localization.}
   \label{fig:fusion}
   \vspace{-0.2in}
\end{figure}

Beyond Wi-Fi, other wireless radio frequency (RF) signals can be used for indoor localization such as ultra-wideband (UWB), Bluetooth, ZigBee, and inertial measurement unit (IMU). Localizing indoor objects was achieved using the IMU with tracking errors adjusted by the Wi-Fi \cite{ChenWu14}. The fusion of Wi-Fi and IMU was later improved by \cite{ChoiChoi20} to predict the trajectory without calibration. DeepFusion combines heterogeneous wearable (e.g., smartphone and smartwatch) and wireless (Wi-Fi and acoustic) sensors for human activity recognition \cite{XueJiang19}. The fusion of the coarse-grained RSSI, Bluetooth, and ZigBee data was considered in \cite{RodriguesVieira11}. In \cite{AhmedArablouei19}, the fine-grained CSI was used to extract the angle and then fused with RSSI from Bluetooth. 

In this paper, we consider multi-modal indoor localization using Wi-Fi, IMU, and UWB. Specifically, we formulate the multi-modal localization as a multi-stream recurrent neural network. Each stream is designed to learn its underlying state evolution to the localization in a supervised learning fashion. Different from other heterogeneous fusion methods for indoor localization, the proposed recurrent fusion network directly learns the modality quality, namely relative importance weights, from its own immediate past states as we conjecture that these immediate past states exhibit highly relevant features to assess the quality of new sensor measurement at the current step. Then the learned importance weights are used to fuse the current states to infer the current location coordinate. 

\vspace{-0.12in}

\begin{figure*}[t]
     \begin{center}
     \includegraphics[width=0.75\textwidth]{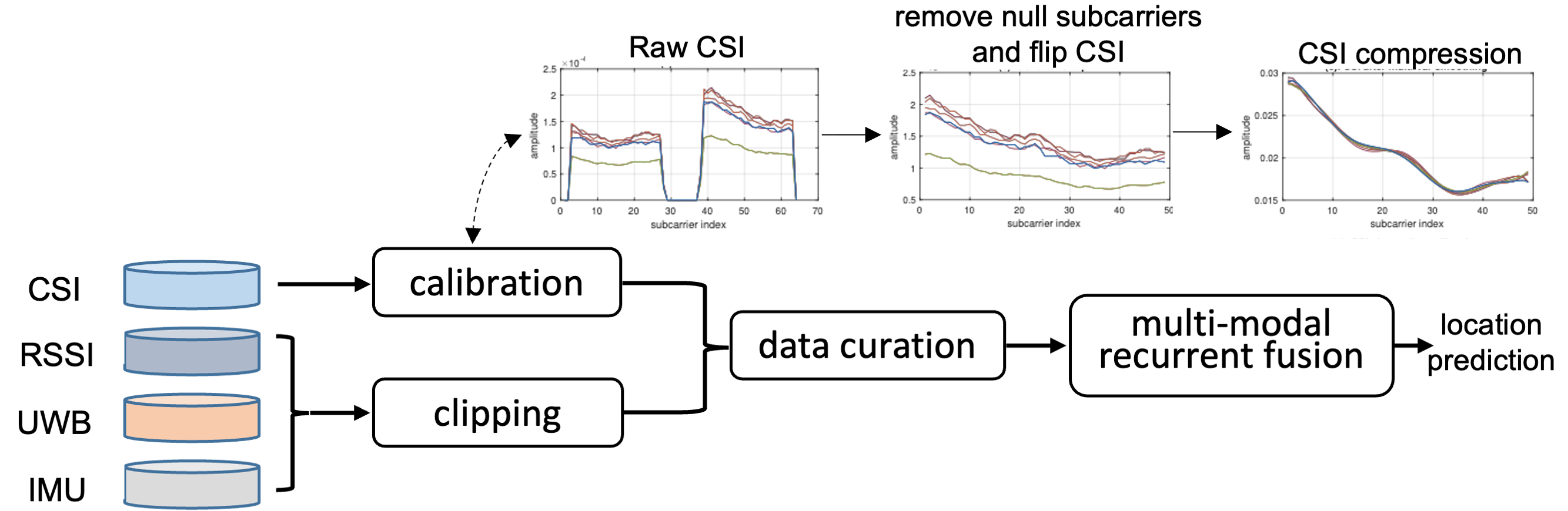}
        \end{center}
    \vspace{-0.2in}
    \caption{The overall multi-modal indoor localization pipeline including the measurement calibration, data curation, and the multi-modal recurrent fusion block.}
  \label{fig:spawc_flowchart}
      \vspace{-0.15in}
\end{figure*}

\section{Problem Formulation}
Like the traditional fingerprinting-based method, we first collect multi-modal wireless sensor measurements at known positions as the fingerprinting dataset. Particularly, we use 
\begin{align}
x^r_i(t), \xbf^{c}_i(t) \in \Cset^{N_c \times 1} & \rightarrow \text{RSSI, CSI of the $i$-th Wi-Fi anchor}  \\
\xbf^{UWB}_i(t) \in \Rset^{2 \times 1} & \rightarrow \text{(range, power) of the $i$-the UWB anchor} \notag \\
\xbf^{IMU}(t) \in \Rset^{9 \times 1} & \rightarrow \text{$3$-axis (acceleration, gyro, and magnetic)}, \notag
\end{align}
where $N_c$ is the number of Wi-Fi subcarriers. 
Assuming $L_w$ and $L_u$ anchors for the Wi-Fi and UWB, the following multi-modal sensor measurements form the fingerprinting data at each time step $t$, 
\begin{align}
\xbf^{MM}_t = \{  \xbf^r_t \in \Rset^{L_w \times 1}, & \ \ \xbf^c_t \in \Cset^{N_c L_w \times 1}, \notag \\
 \xbf^{UWB}_t \in \Rset^{2L_u \times 1}, & \ \ \xbf^{IMU}_t \in \Rset^{9 \times 1}  \}, 
\end{align}

For each $t$, the two-dimensional coordinate $\{x_t, y_t\}$ is known. The multi-modal localization is to infer the coordinate from the multi-sensor data over a certain time interval:
\begin{align}
\{\xbf^{MM}_i \}_{i=t-T+1}^t \rightarrow \{x_t, y_t \},
\end{align}
where $T$ is the number of time steps used to estimate the coordinate. 

\section{Multi-Stream Recurrent Fusion Network for Indoor Localization}

The overall pipeline is shown in Fig.~\ref{fig:spawc_flowchart} with the multi-model recurrent fusion block given in Fig.~\ref{fig:fusion}. In Fig.~\ref{fig:fusion}, each sensor stream takes a sequence of one type of sensor measurements and uses the recurrent neural units (e.g., long short-term memory (LSTM)) to learn the long-term hidden state evolution. Then, multiple hidden states immediately preceding the current step are projected to relative importance weights, a measure of relative data quality for each sensor. The hidden states at the current step are combined by these importance weights to predict the coordinate. 

\subsection{Data Curation}

Commodity CSI measurements are known to be impacted by hardware impairments such as the sample frequency offset (SFO) and carrier frequency offset (CFO). 
To mitigate these impacts, we take a standard procedure to compress the raw CSI data. As shown in the top figures of Fig.~2, null subcarriers are first removed and the remaining ones are flipped to smooth the whole frequency spectrum. The flipped CSI is locally calibrated by normalizing each CSI as $\tilde{\xbf}^c_i(t) = |\xbf^c_i(t)|  / \sum_j | x^c_{i,k}(t) |$ where $x^c_{i,k}(t)$ is the $k$-th subcarrier CSI of the $i$-th anchor at time $t$, which fixes the gain fluctuation due to the use of automatic gain control (AGC) at commodity Wi-Fi devices. To reduce the data overhead, we apply the polynomial fitting to compress the calibrated CSI into a vector $\abf_i(t)$~\cite{SobehyRenault19}, 
\begin{align}
\tilde{\xbf}^c_i(t) = \Kbf \abf_i(t)
\end{align} 
where $\Kbf= [\mathbf{1}, \kbf, \cdots, \kbf^P]$ is the polynomial basis matrix with $\kbf$ grouping the remaining subcarrier indices and $P$ denoting the polynomial order, and $\abf_i(t)=[a_{0, i}(t), \cdots, a_{P, i}(t)]^T$ is the coefficient vector for the $i$-th calibrated CSI amplitude. 

For other sensor measurements such as RSSI, IMU, and UWB, we apply a simple clipping step to exclude outliers. For instance, the range measurements from the UWB sensors are clipped at $10$ m, while the RSSI is clipped at $0$ dBm. As a result, we used the following multi-sensor data for indoor localization
\begin{align}
\{ \bar\xbf^r_i, \abf_i, \bar{\xbf}^{UWB}_i, \bar{\xbf}^{IMU}_i \}_{i=t-T+1}^t \rightarrow \{x_t, y_t \}
\end{align}
where $\abf_t=[\abf^T_1(t), \cdots, \abf^T_{L_w}]^T$ groups all CSI coefficient vectors from the $L_w$ anchors, and  $\bar{\xbf}_t$ represent the clipped version of $\xbf_t$.


\subsection{Recurrent Neural Units}
To utilize historic data for indoor localization, we formulate the indoor localization as a sequence regression problem where the current coordinate is estimated from the multi-modal data from the current and previous  time steps. With this sequence-to-coordinate formulation, the RNN is used to learn time-dependent features from the input sequence. Here, we consider the LSTM network \cite{HochreiterSchmidhuber97}. 

In the context of multi-modal indoor localization, an LSTM is to estimate the conditional probability 
\begin{align}
p(x_t, y_t| \{ \bar\xbf^r_i, \abf_i, \bar{\xbf}^{UWB}_i, \bar{\xbf}^{IMU}_i \}_{i=t-T+1}^t).
\end{align} 
Take the RSSI $\bar{\xbf}^r_t$ as an example. The RSSI-stream LSTM can be described with the following process:
\begin{align}
\cbf_{t} &=\fbf_{t} \odot \cbf_{t-1}+\ibf_{t} \odot \tilde{\cbf}_t,  \label{cell} \\
\tilde{\cbf}_t &= \tanh \left(\Wbf_{r c} \bar{\xbf}^r_{t}+\Wbf_{h c} \hbf_{t-1}+\bbf_{c}\right),  \label{candidate} \\
\fbf_{t} &=\sigma\left(\Wbf_{rf} \bar{\xbf}^r_{t}+\Wbf_{h f} \hbf_{t-1}+\bbf_{f}\right), \label{forget} \\
\ibf_{t} &=\sigma \left(\Wbf_{ri}\bar{\xbf}^r_{t}+\Wbf_{hi} \hbf_{t-1}+\bbf_{i}\right), \label{input} \\
\hbf_{t} &=\obf_{t} \odot \tanh \left(\cbf_{t}\right) \label{heq}, \\
\obf_{t} &=\sigma\left(\Wbf_{r o} \bar{\xbf}^r_{t}+\Wbf_{h o} \hbf_{t-1}+\Wbf_{c o} \odot \cbf_{t}+ \bbf_{o}\right) \label{output}, 
\end{align}
Specifically, in \eqref{cell}, a memory cell content $\cbf_t$ updates its ``old" memory content $\cbf_{t-1}$ passing through the ``current" forget gate output $\fbf_t$ and adds new ``candidate" memory cell $\tilde{\cbf}_t$ weighted by the ``current" input gate output $\ibf_t$.  The candidate memory cell, computed in \eqref{candidate}, uses the tanh function to combine the previous hidden state $\hbf_{t-1}$ and the current input $\bar{\xbf}^r_{t}$, plus a bias term $\bbf_c$, into a value range of $(-1, 1)$. The forget gate of \eqref{forget} also acts on $(\hbf_{t-1}, \bar{\xbf}^r_{t})$ but compresses the value into $(0, 1)$ with the sigmoid function $\sigma(\cdot)$ to determine how much of the old memory cell content $\cbf_{t-1}$ should retain in \eqref{cell}. Similarly, the input gate of \eqref{input} compresses $(\hbf_{t-1},\bar{\xbf}^r_{t})$ into another value in between $0$ and $1$ and decides how much information we should take from the new input $\bar{\xbf}^r_{t}$ via $\tilde{\cbf}_f$  in \eqref{cell}. Finally, the new hidden state $\hbf_t$ is updated in \eqref{heq} with $\obf_t$ as the output gate in \eqref{output}. 

In other words, the new hidden state is a gated version (via $\obf_t$) of the tanh of the memory cell $\cbf_t$. For instance, when the output gate is close to $1$, the hidden state is effectively the memory cell content to the next time step or the regression head. One can stack multiple LSTM layers and use not only the forward pass but also the backward pass for a bi-directional choice \cite{SinghMarks16}. 

\subsection{Relative Importance Weights}
Compared with the standard multi-stream LSTM \cite{SinghMarks16} for computer vision applications, the data quality of multi-modal RF sensors may vary drastically over time due to the surrounding environment, e.g., non-line-of-sight (NLOS) scenarios, and sensor failures. For instance, the UWB works well when the user is in sight while degrading quickly at NLOS locations. Motivated by the observation, we propose to project a concatenated ``immediately preceding" hidden states $\hbf_u = [\hbf^T_{t-F}, \cdots, \hbf^T_{t-1}]^T$ into a measure of modality quality, $u$, and normalize these quality measures to reflect the relative importance of each sensor at time $t$. Specifically, we utilize a fully-connected (FC) layer followed by the sigmoid  function:
\begin{align}
u = \sigma \left(\wbf^T_{u} \hbf_u + b_u \right)
\end{align}
where  $F$ is the number of past hidden states, $\wbf_u$ is the corresponding FC coefficient vector of dimension $F N_h \times 1$ with $N_h$ denoting the hidden state dimension, and $b_u$ is the bias term. Collecting all $u$ for selected sensor types, we use the softmax function to normalize them to the relative importance weights
\begin{align}
\alpha_m = \frac{e^{u_m}}{\sum_{m=1}^M e^{u_m}} \in [0, 1],
\end{align}
where $m=1, \cdots, M$ refers to the sensor type such as the RSSI, CSI, IMU and UWB, and $\sum_m \alpha_m = 1$. 

\subsection{Recurrent Fusion}
With the multi-stream LSTM and the importance weight estimates, we propose to fuse the last hidden states from multiple LSTM streams, weighted by their relative importance:
\begin{align}
\hbf_{\text{fusion}}= \sum_{m=1}^M \alpha_m \hbf^m_t.
\end{align}
The fused state is then fed into the coordinate estimation block which consists of several FC layers along with the ReLU activation $\sigma_{\text{R}}(\cdot)$:
\begin{align}
[\hat{x}_t; \hat{y}_t]= \Wbf^Q_{o} \vbf^Q_o + \bbf^Q_o,  \vbf^q_o = \sigma_{\text{R}}( \Wbf^{q-1}_{o} \vbf^{q-1}_o + \bbf^{q-1}_o),
\end{align}
where $q=1, \cdots,Q$, $\vbf^0_o = \hbf_{\text{fusion}}$, and $Q$ is the number of FC layers. We train the multi-modal fusion network in an end-to-end fashion with the loss function of mean squared error (MSE) between the estimated coordinate $[\hat{x}_t, \hat{y}_t]$ and the ground truth $[{x}_t, {y}_t]$. 

\emph{Remark}: Kalman filter-like approaches also use relative importance or uncertainty for multi-sensor fusion via the propagation of (cross- and self-) covariance matrices of multiple sensor modalities with known measurement and (and likely Markovian) dynamics models. In contrast, the multi-stream LSTM is a data-driven approach that utilizes a standard LSTM to learn both nonlinear measurement and dynamics models over a long-term horizon and introduces a nonlinear mapping of immediate past hidden states to estimate the relative importance.

%


\begin{figure}[t]
     \begin{center}
     \includegraphics[width=0.45\textwidth]{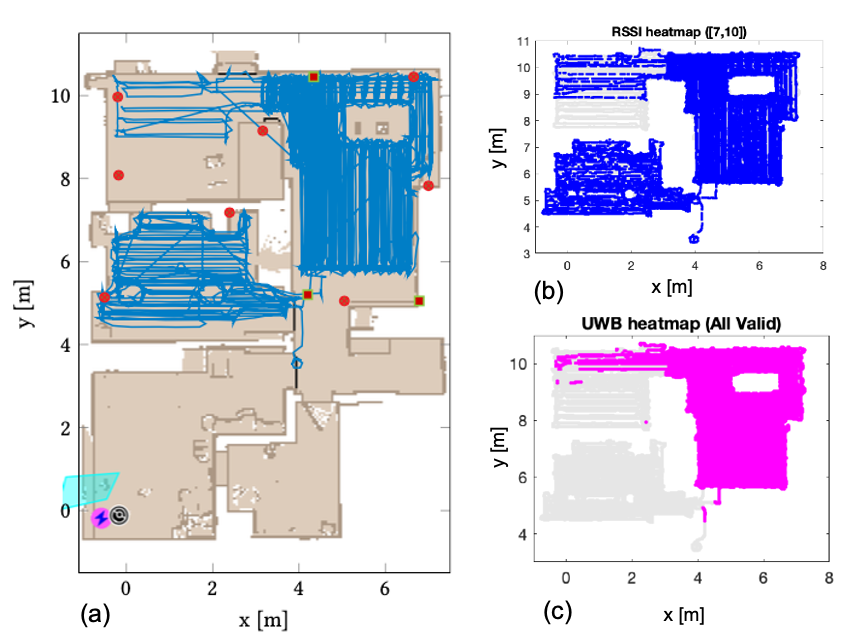}
        \end{center}
    \vspace{-0.2in}
    \caption{The floor plan of the \emph{SPAWC2021 multi-modal localization dataset} \cite{ArnoldSchaich21} where red squares and  red circles denote, respectively, the UWB and Wi-Fi anchor locations. }
   \label{fig:floorplan}
  \vspace{-0.1in}
\end{figure}

\begin{figure}
 \centering
  \subfloat[Individual sensor]{\includegraphics[width=0.33\textwidth]{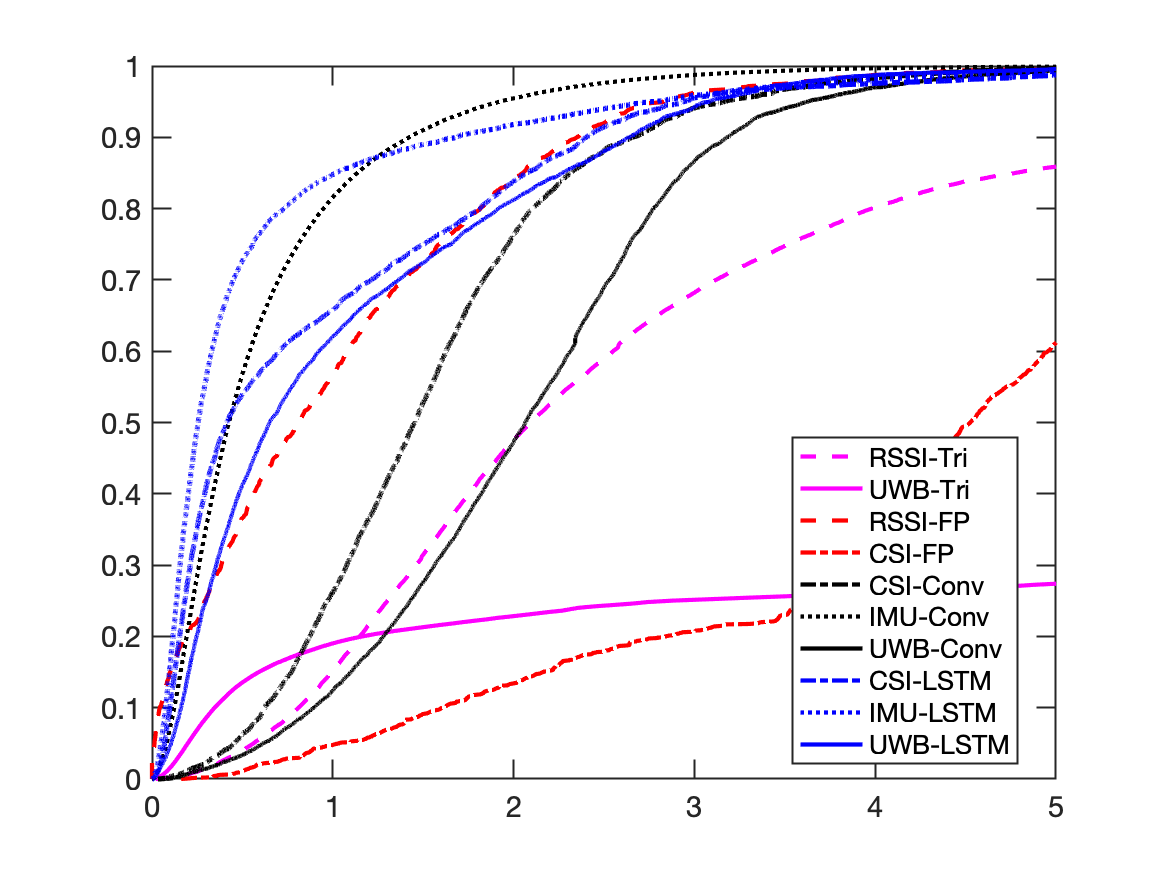}} \hfil
 \subfloat[Sensor fusion]{\includegraphics[width=0.33\textwidth]{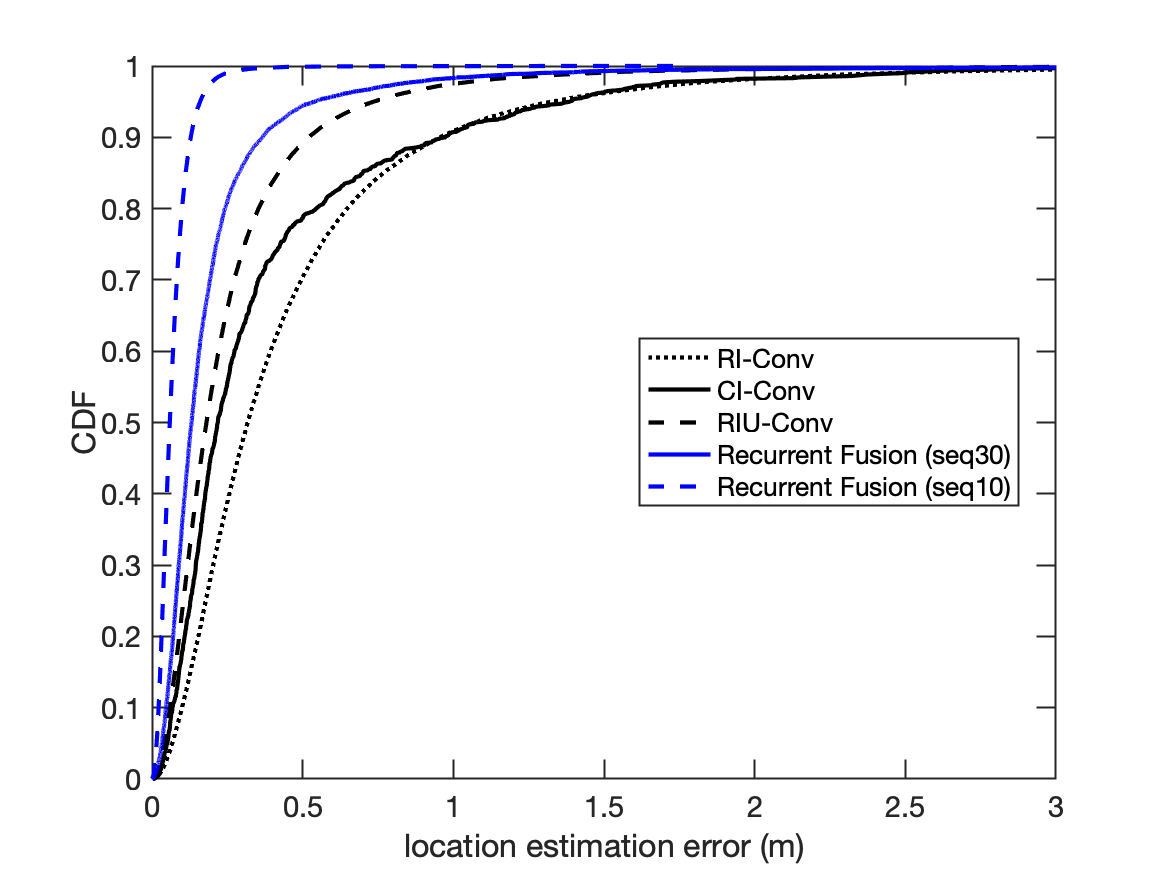}}
\caption{The cumulative distribution function (CDF) of localization errors for the recurrent fusion methods and a list of baseline methods. }
\label{fig_coorEst}	
\end{figure}

\section{Performance Evaluation}
\label{sec:performance}
In the following, we present experimental evaluation on the \textit{SPAWC2021 multi-modal localization} dataset. 

\subsection{SPAWC2021 Multi-Modal Localization Dataset}
The SPAWC2021 multi-modal localization dataset was collected using a robot in a multi-room setting with a square footage about $8 \text{ m} \times 8 \text{ m}$, as shown in Fig.~\ref{fig:floorplan} (a). The robot platform was mounted with a multi-sensor testbed consisting of multiple commercial-of-the-shelf Wi-Fi (ESP$32$), UWB (DW$1000$), and IMU (BNO$055$) chipsets  \cite{ArnoldSchaich21}. The pseudo-ground-truth coordinates were obtained from the SLAM algorithm from a two-dimensional Lidar, ultrasonic distance sensors, wheel-tick sensors, infrared sensors, and the IMU. More details about the dataset can be found in \cite{ArnoldSchaich21}.

We consider \emph{Dataset 1} with about $750k$ samples of the three sensor data. The covered robot trajectories are highlighted in blue in Fig.~\ref{fig:floorplan} (a) for a course of $1975.28$ m. For each sampled location, RSSI (scalar) and CSI ($64$ subcarriers) from $11$ Wi-Fi anchors, range and power readings from $3$ UWB anchors, and a $9$-dimensional IMU data were recorded. Fig.~\ref{fig:floorplan} (b) and (c) show the RSSI and UWB heatmaps, respectively, when $7\mathrm{-}10$ Wi-Fi and all $3$ UWB anchors report valid readings. It is seen that the  upper-right room is well covered by all $3$ UWB anchors, while the lower-left room has no UWB coverage due to the NLOS configuration. 

\subsection{Implementation}
\label{sec:implementation}
We standardize each input entry by subtracting the mean and normalizing it with the standard deviation. For the LSTM, we use a network configuration of $2$ layers, a hidden dimension of $N_h=256$, an optional bi-directional implementation, an uncertainty estimation block of a varying number of hidden states $F=\{1, 2\}$, and an output block of $Q=2$ FC layers with a hidden layer dimension of $128$. 
To regularize the training from overfitting, we adopt the dropout that randomly sets the hidden state outputs of both the first and second LSTM layers to zeros with a probability of $0.2$\cite{SrivastavaHinton14}. 

We consider various combinations of sensor fusion, ranging from $M=1$ (no fusion) to $M=3$ (three types of sensors). Two sequence lengths $T=10$ and $T=30$ with corresponding stepsizes of $1$ and $2$, respectively. We train the network using the Adam optimizer with a weight decay of $1\mathrm{e}{-4}$ and a learning rate of $1\mathrm{e}{-3}$. For up to $50 \times 10^4$ iterations, the models are trained with a mini-batch size of $128$ by an $80/10/10$ splitting of Dataset1 among the training, validation, and testing sets, as the same as \cite{ArnoldSchaich21}. 

\subsection{Comparison to Baseline Methods}
We include the following baseline methods: 1) two trilateration methods (\textit{X}-Tri with \textit{X} denoting the sensor type) \cite{HeChan16}; 2) two $k$-nearest neighbor methods (\textit{X}-FP) that idenfity $k=2$ nearest fingerprinting samples to the input and apply a linear interpolation between the two fingerprinted locations \cite{ZafariGkelias19}; 3) three single-frame convolution neural network method (\textit{X}-Conv) \cite{ChenZhang17} with the architecture auto-tuned by the AutoML tool \cite{ArnoldSchaich21}; and 4) three convolution-based fusion methods \cite{ArnoldSchaich21}. We also include four individual-stream LSTM (\textit{X}-LSTM) of $M=1$. The performance metric is the cumulative distribution function (CDF) of localization errors. 

Fig.\ref{fig_coorEst} (a) compares the trilateration method, fingerprinting methods, convolution net-based methods, and the individual stream LSTM methods. First, as observed in \cite{ArnoldSchaich21}, the trilateration (magenta) and fingerprinting (red) methods give more large localization errors, e.g., more than $3$ m, partly due to the presence of NLOS areas. Second, except for the CSI data, the \textit{X}-Conv methods (black) show better performance. Particularly, the IMU-Conv method provides the best result among the \textit{X}-Conv methods as the IMU data is less impacted by the NLOS areas. Third, the sequential \textit{X}-LSTM methods (blue) lead to further improved localization performance. This improvement is noticeable for the UWB sensor by comparing the three solid curves. 

Fig.\ref{fig_coorEst} (b) compares the convolution-based fusion methods, i.e., the RI-Conv, CI-Conv, and RIU-Conv, with the proposed recurrent fusion method. For these two considered sequence lengths $T=10$ and $T=30$, the performance is better than the RIU-Conv result, the best one of the convolution-based fusion class \cite{ArnoldSchaich21}. Table~\ref{tab:results} further summarizes quantitative performance in terms of the mean, median, and the location error corresponding to the $90$th percentile of the CDF. For the recurrent fusion (RF) method, we list these error metrics with various configurations. 
Several observations can be made here. First, $F=1$ appears to yield better results than $F=2$. This is potentially due to the redundant information included in the later hidden state and more parameters to be learned with a larger $F$. Second, the bi-directional option shows only marginal performance improvements when $F=1$. Finally, a short sequence of $T=10$ leads to a bigger performance gain than other configuration parameters. The best RF version shows a performance improvement of an order of magnitude over the RIU-Conv method.

\begin{table}
  \caption{Localization errors (m) on \textit{SPAWC2021} dataset.}\vspace{-3mm}
  \label{tab:results}
  \centering
  \setlength\tabcolsep{4pt}
  {
  \begin{tabular}{lccccccc}
    \toprule
  
    & Mean & Median & CDF@0.9  \\
    
    \midrule
    
    RSSI-Tri & 13.74 & 2.09 & 7.00 & \\
    UWB-Tri & 7.57 & 9.36 & 11.54 & \\
    RSSI-FP & 1.05 & 0.81 & 2.35 & \\
    CSI-FP & 4.18 & 4.53 & 6.32 & \\
    CSI-Conv & 1.53 & 1.45 & 2.66 & \\
    UWB-Conv & 1.99 & 2.07 & 3.18 & \\
    IMU-Conv  & \textbf{0.64} &  \textbf{0.43} &  \textbf{1.42} & \\
    
    \midrule
    
    RSSI-LSTM & 1.10 & 0.75 & 2.43 & \\
    CSI-LSTM & 0.94 & 0.41 & 2.40 & \\
     UWB-LSTM & 1.07 & 0.65 & 2.64 & \\
     IMU-LSTM & \textbf{0.60} & \textbf{0.26} & \textbf{1.68} & \\
     
     \midrule

    RI-Conv & 0.46 & 0.32 & 0.96 & \\
    CI-Conv & 0.39 & 0.22 & 0.97 & \\
    RIU-Conv & \textbf{0.26} & \textbf{0.18} & \textbf{0.52} & \\   
    
     \midrule
    
    RF(F=1, seq30, biTrue) & 0.18 & 0.12 & 0.33 & \\
    RF(F=1, seq30, biFalse) & 0.19 & 0.13 & 0.37 & \\
    RF(F=1, seq10, biTrue) & \textbf{0.06} & \textbf{0.05} & \textbf{0.12} & \\
    RF(F=1, seq10, biFalse) & 0.07 & 0.06 & 0.13 & \\
    
    RF(F=2, seq30, biTrue) & 0.33 & 0.18 & 0.63 & \\
    RF(F=2, seq30, biFalse) & 0.19 & 0.12 & 0.35 & \\
    
    
        \bottomrule
  \end{tabular}
  }
\end{table}

\section{Conclusion}
This paper has considered the multi-modal sensor fusion for indoor localization in a sequential learning setting. Specifically, we proposed to fuse the hidden states from these multi-modal sensors, weighted by their relative importance directly learned from their past hidden states. Comprehensive performance comparison on a large-scale open dataset confirms significant performance gain of the proposed method over a list of baseline methods.


\section{Acknowledgement}
The authors sincerely thank Dr.~Maximilian Arnold from Nokia Bell Labs (Germany) for releasing the \textit{SPAWC2021 multi-modal localization} dataset and providing the original figure data in \cite{ArnoldSchaich21} for the baseline comparison.

\newpage

\bibliographystyle{IEEEbib}
\bibliography{bib_localization}

\begin{thebibliography}{10}

\bibitem{HeChan16}
S.~{He} and S.~H. {Chan},
\newblock ``{Wi-Fi} fingerprint-based indoor positioning: Recent advances and
  comparisons,''
\newblock {\em IEEE Communications Surveys Tutorials}, vol. 18, no. 1, pp.
  466--490, 2016.

\bibitem{ZafariGkelias19}
F.~Zafari, A.~Gkelias, and K.K. Leung,
\newblock ``A survey of indoor localization systems and technologies,''
\newblock {\em IEEE Communications Surveys \& Tutorials}, vol. 21, no. 3, pp.
  2568--2599, 2019.

\bibitem{BahlPadmanabhan00}
P.~{Bahl} and V.~N. {Padmanabhan},
\newblock ``{RADAR}: an in-building {RF-based} user location and tracking
  system,''
\newblock in {\em INFOCOM}, March 2000, vol.~2, pp. 775--784.

\bibitem{WuXiao13}
K.~{Wu} {et. al.},
\newblock ``{CSI}-based indoor localization,''
\newblock {\em IEEE Trans. on Parallel and Distributed Systems}, vol. 24, no.
  7, pp. 1300--1309, July 2013.

\bibitem{WangGao15wcnc}
X.~{Wang}, L.~{Gao}, S.~{Mao}, and S.~{Pandey},
\newblock ``{DeepFi}: Deep learning for indoor fingerprinting using channel
  state information,''
\newblock in {\em WCNC}, March 2015, pp. 1666--1671.

\bibitem{WangGao17}
X.~{Wang}, L.~{Gao}, S.~{Mao}, and S.~{Pandey},
\newblock ``{CSI}-based fingerprinting for indoor localization: A deep learning
  approach,''
\newblock {\em IEEE Transactions on Vehicular Technology}, vol. 66, no. 1, pp.
  763--776, Jan 2017.

\bibitem{ChenZhang17}
H.~{Chen}, Y.~{Zhang}, W.~{Li}, X.~{Tao}, and P.~{Zhang},
\newblock ``{ConFi}: Convolutional neural networks based indoor {Wi-Fi}
  localization using channel state information,''
\newblock {\em IEEE Access}, vol. 5, pp. 18066--18074, 2017.

\bibitem{PajovicWang19}
M.~Pajovic, P.~Wang, T.~Koike-Akino, H.~Sun, and P.~V. Orlik,
\newblock ``Fingerprinting-based indoor localization with commercial {MMWave
  WiFi}--{Part I: RSS and Beam Indices},''
\newblock in {\em GLOBECOM}, Dec. 2019.

\bibitem{WangPajovic19}
P.~Wang, M.~Pajovic, T.~{Koike-Akino}, H.~Sun, and P.V. Orlik,
\newblock ``Fingerprinting-based indoor localization with commercial mmwave
  {WiFi}--{Part II}: Spatial beam {SNRs},''
\newblock in {\em GLOBECOM}, Dec 2019.

\bibitem{KoikeWang20}
T.~{Koike-Akino}, P.~{Wang}, M.~{Pajovic}, H.~{Sun}, and P.~V. {Orlik},
\newblock ``Fingerprinting-based indoor localization with commercial mmwave
  {WiFi}: A deep learning approach,''
\newblock {\em IEEE Access}, vol. 8, pp. 84879--84892, 2020.

\bibitem{WangKoike20b}
P.~{Wang}, T.~{Koike-Akino}, and P.~V. {Orlik},
\newblock ``Fingerprinting-based indoor localization with commercial mmwave
  {WiFi}: {NLOS} propagation,''
\newblock in {\em GLOBECOM}, December 2020.

\bibitem{YuWang20}
J.~Yu, P.~{Wang}, T.~{Koike-Akino}, and P.~V. {Orlik},
\newblock ``Human pose and seat occupancy classification with commercial mmwave
  {WiFi},''
\newblock in {\em GLOBECOM Workshop on Integrated Sensing and Communication
  (ISAC)}, December 2020.

\bibitem{YuWang22}
J.~Yu, P.~Wang, T.~Koike-Akino, Y.~Wang, P.~V. Orlik, and R.~M. Buehrer,
\newblock ``Multi-band {Wi-Fi} sensing with matched feature granularity,''
\newblock {\em arXiv:2112.14006}, 2022.

\bibitem{KoikeWang22}
T.~{Koike-Akino}, P.~{Wang}, and Y.~Wang,
\newblock ``Quantum transfer learning for {Wi-Fi} sensing,''
\newblock in {\em IEEE International Conference on Communications (ICC)}, May
  2022.

\bibitem{YoussefAgrawala08}
M.~Youssef and A.~Agrawala,
\newblock ``The horus location determination system,''
\newblock {\em Wirel. Netw.}, vol. 14, no. 3, pp. 357--374, June 2008.

\bibitem{MazuelasBahillo09}
S.~{Mazuelas} {et. al.},
\newblock ``Robust indoor positioning provided by real-time {RSSI} values in
  unmodified {WLAN} networks,''
\newblock {\em IEEE Journal of Selected Topics in Signal Processing}, vol. 3,
  no. 5, pp. 821--831, Oct 2009.

\bibitem{LiZhang14}
D.~{Li}, B.~{Zhang}, Z.~{Yao}, and C.~{Li},
\newblock ``A feature scaling based k-nearest neighbor algorithm for indoor
  positioning system,''
\newblock in {\em GLOBECOM}, Dec 2014, pp. 436--441.

\bibitem{WangGao16}
X.~{Wang}, L.~{Gao}, and S.~{Mao},
\newblock ``{CSI} phase fingerprinting for indoor localization with a deep
  learning approach,''
\newblock {\em IEEE Internet of Things Journal}, vol. 3, no. 6, pp. 1113--1123,
  Dec 2016.

\bibitem{WangGao17b}
X.~{Wang}, L.~{Gao}, and S.~{Mao},
\newblock ``Biloc: Bi-modal deep learning for indoor localization with
  commodity {5GHz WiFi},''
\newblock {\em IEEE Access}, vol. 5, pp. 4209--4220, 2017.

\bibitem{HsiehChen19}
C.~{Hsieh}, J.~{Chen}, and B.~{Nien},
\newblock ``Deep learning-based indoor localization using received signal
  strength and channel state information,''
\newblock {\em IEEE Access}, vol. 7, pp. 33256--33267, 2019.

\bibitem{HoangYuen19}
M.~T. {Hoang}, B.~{Yuen}, X.~{Dong}, T.~{Lu}, R.~{Westendorp}, and K.~{Reddy},
\newblock ``Recurrent neural networks for accurate {RSSI} indoor
  localization,''
\newblock {\em IEEE Internet of Things Journal}, vol. 6, no. 6, pp.
  10639--10651, Dec 2019.

\bibitem{WuLi07}
Z.~{Wu} {et. al.},
\newblock ``Location estimation via support vector regression,''
\newblock {\em IEEE Trans. on Mobile Computing}, vol. 6, no. 3, pp. 311--321,
  March 2007.

\bibitem{ChenWu14}
L.~{Chen}, E.~H. {Wu}, M.~{Jin}, and G.~{Chen},
\newblock ``Intelligent fusion of {Wi-Fi} and inertial sensor-based positioning
  systems for indoor pedestrian navigation,''
\newblock {\em IEEE Sensors Journal}, vol. 14, no. 11, pp. 4034--4042, 2014.

\bibitem{ChoiChoi20}
J.~Choi and Y.-S. Choi,
\newblock ``Calibration-free positioning technique using {Wi-Fi} ranging and
  built-in sensors of mobile devices,''
\newblock {\em IEEE Internet of Things Journal}, vol. 8, no. 1, pp. 541--554,
  2020.

\bibitem{XueJiang19}
H.~Xue et~al.,
\newblock ``{DeepFusion}: A deep learning framework for the fusion of
  heterogeneous sensory data,''
\newblock in {\em MobiHoc}, 2019, pp. 151--160.

\bibitem{RodriguesVieira11}
M.~L. {Rodrigues}, L.~F.~M. {Vieira}, and M.~F.~M. {Campos},
\newblock ``Fingerprinting-based radio localization in indoor environments
  using multiple wireless technologies,''
\newblock in {\em PIMRC}, 2011, pp. 1203--1207.

\bibitem{AhmedArablouei19}
A.~U.~Ahmed {et. al.},
\newblock ``Multi-radio data fusion for indoor localization using {Bluetooth}
  and {WiFi},''
\newblock in {\em PECCS}, 2019, pp. 13--24.

\bibitem{SobehyRenault19}
A.~Sobehy, E.~Renault, and P.~Muhlethaler,
\newblock ``{CSI} based indoor localization using ensemble neural networks,''
\newblock {\em IFIP MLN}, pp. 367--378, 2019.

\bibitem{HochreiterSchmidhuber97}
S.~Hochreiter and J.~Schmidhuber,
\newblock ``{Long Short-Term Memory},''
\newblock {\em Neural Computation}, vol. 9, no. 8, pp. 1735--1780, 11 1997.

\bibitem{SinghMarks16}
B.~Singh et~al.,
\newblock ``A multi-stream bi-directional recurrent neural network for
  fine-grained action detection,''
\newblock in {\em CVPR}, 2016, pp. 1961--1970.

\bibitem{ArnoldSchaich21}
M.~Arnold and F.~Schaich,
\newblock ``Indoor positioning systems: Smart fusion of a variety of sensor
  readings,''
\newblock {\em arXiv:2105.05438}, 2021.

\bibitem{SrivastavaHinton14}
N.~Srivastava et~al.,
\newblock ``Dropout: A simple way to prevent neural networks from
  overfitting,''
\newblock {\em Journal of Machine Learning Research}, vol. 15, no. 56, pp.
  1929--1958, 2014.

\end{thebibliography}

\end{document}